\documentclass[aps,prb,twocolumn,superscriptaddress,floatfix,showpacs]{revtex4}

\usepackage{graphicx}
\begin{document}

\title{Superconductivity of Mg/MgO interface formed by shock-wave pressure}

\author{N.~S.~Sidorov}
\author{A.~V.~Palnichenko}
\affiliation{Institute of Solid State Physics, Russian Academy of Sciences,
Chernogolovka, Moscow region, 142432, Russia}

\author{D.~V.~Shakhrai}
\author{V.~V.~Avdonin}
\affiliation{Institute of Problems of Chemical Physics, Russian Academy of
Sciences, Chernogolovka, Moscow region, 142432, Russia}

\author{O.~M.~Vyaselev}
\author{S.~S.~Khasanov}
\affiliation{Institute of Solid State Physics, Russian Academy of Sciences,
Chernogolovka, Moscow region, 142432, Russia}

\date{\today}

\begin{abstract}

A mixture of Mg and MgO has been subjected to a shock-wave pressure of
$\simeq\,$170\,kbar. The ac susceptibility measurements of the product has
revealed a metastable superconductivity with $T_c \approx 30\,$K, characterized
by glassy dynamics of the shielding currents below $T_c$. Comparison of the ac
susceptibility and the dc magnetization measurements infers that the
superconductivity arises within the interfacial layer formed between metallic
Mg and its oxide due to the shock-wave treatment.

\end{abstract}


\maketitle

\section{Introduction}

Interfacial superconductivity is one of the emergent phenomena that appears at
the boundary between some materials, even insulators, as a consequence of a
large change of physical properties of these materials at the interface,
compared to their properties in the bulk.\cite{Hwang, Pereiro, Gariglio} It has
been experimentally evidenced since the 1950's and 60's that specifically
fabricated thin layers of some superconductors (e.g. sputtered onto a cold
substrate of sandwiched between non-superconducting materials) demonstrate a
higher superconducting transition temperature, $T_c$, compared to that in the
bulk.\cite{Zavaritsky, Buckel, PRL67, PRL68, JAP68, Hauser} However, it took
several decades before the necessary technologies providing wide abilities of
fabricating two-dimensional heterostructures controlled on the molecular and
atomic scale were developed, which enabled the creation of many novel materials
on the basis of interfaces.\cite{Rijnders,Logvenov}

The superconducting interfaces known currently, belong to the families of
semiconductor/semiconductor interfaces based on different
chalcogenides,\cite{Murase,Mironov,StronginJS,Fogel} insulator/insulator
interfaces based on LaAlO$_3$/SrTiO$_3$ and related
systems,\cite{Reyren07,Reyren09} as well as metal/insulator interfaces based on
complex oxides such as La$_{1.55}$Sr$_{0.45}$CuO$_4$/La$_2$CuO$_4$.\cite{Gozar}
The origin of the interfacial superconductivity is not completely understood
yet. Moreover, there are only few examples of the interfacial superconducting
layer with $T_c$ higher than in the bulk optimally doped samples.
\cite{Pereiro} The complexity of the used oxides makes the issue even more
obscure. Therefore, discovery of superconducting interfaces formed by simplest
non-superconducting substances could illuminate the issue by revealing the key
factors responsible for the interface superconductivity.

Mg/MgO system formed by thermal surface oxidation of bulk metallic Mg samples,
is one of the first of such elementary systems where the interface between
non-superconducting constituents has been reported to superconduct at
39-54\,K.\cite{SidSSS12,SidPC12} Historically, this experiment was motivated by
earlier efforts to increase $T_c = 39\,$K in magnesium diboride, MgB$_2$, by
heating it in the presence of alkali metals. The efforts failed for pure
MgB$_2$, but resulted in $T_c$ up to 45-58\,K for the mixture of MgB$_2$ and
MgO,\cite{Paln} which was attributed to metal/metal oxide interfaces formed by
alkali metal chemical reduction of MgO inclusions. The above experiments have
shown also an instability of the superconducting Mg/MgO-interface at room
temperature. Superconductivity of a Mg/MgO sample, vacuum-encapsulated in
quartz ampoule, decays gradually during its exposure to room temperature, while
storing the ampoule in liquid nitrogen prevents the sample degradation
infinitely long. This suggests the importance to quench the non-equilibrium
sample to low temperature in order to stabilize the superconducting
Mg/MgO-interface.\cite{SidSSS12,SidPC12}

Such instability of the non-equilibrium superconducting Mg/MgO-interface under
normal conditions has motivated our attempt to create it using shock-wave
pressure. During the shock-wave impact, a stroke applied to the sample creates
a series of strong high-pressure shock-waves propagating throughout the sample
due to relative displacements of local parts of the sample
material.\cite{Zeldovich,Kanel,Fortov} Highly non-equilibrium conditions thus
realized, can stimulate phase transitions or mechanochemical reactions
inaccessible in a static pressure mode.\cite{Ossipyan,150SSC2010} Furthermore,
the energy of the shock wave rapidly propagating through the sample within
10$^{-6}$ -- 10$^{-9}$\,s,\cite{Zeldovich,Kanel,Fortov,Ossipyan} leads to local
non-equilibrium overheat of the sample's regions at the shock wavefront,
followed by their rapid cooling (quenching) as the shock-wave is passed. Such
quenching can provide room-temperature stabilization of metastable
non-equilibrium phases, unstable otherwise under normal conditions.

In this paper we report on metastable superconductivity at $\approx 30$\,K
revealed by the ac magnetic susceptibility measurements of the mixture of Mg
and MgO subjected to shock-wave pressure of $\simeq\,170$\,kbar.

\section{Experimental}

\subsection{Sample preparation}

The samples were prepared using flat-type shock-wave pressure setup, machined
of X18H10 stainless steel, described earlier in Refs.~\onlinecite{Ossipyan,
150SSC2010}. The starting samples were tablets, 9.6\,mm in diameter and 0.9\,mm
thick, prepared of 99.99\,\%-pure metallic magnesium covered by 0.2\,mm layer
of powdered, 10 - 50\,$\mu$m grain size, 99.99\,\%-pure magnesium oxide, MgO.
Prior to the shock-wave pressure treatment, the samples were compacted to
reduce porosity using static pressure up to 0.3\,kbar.

\begin{figure}[t]
\includegraphics[width=0.7\linewidth, angle=90]{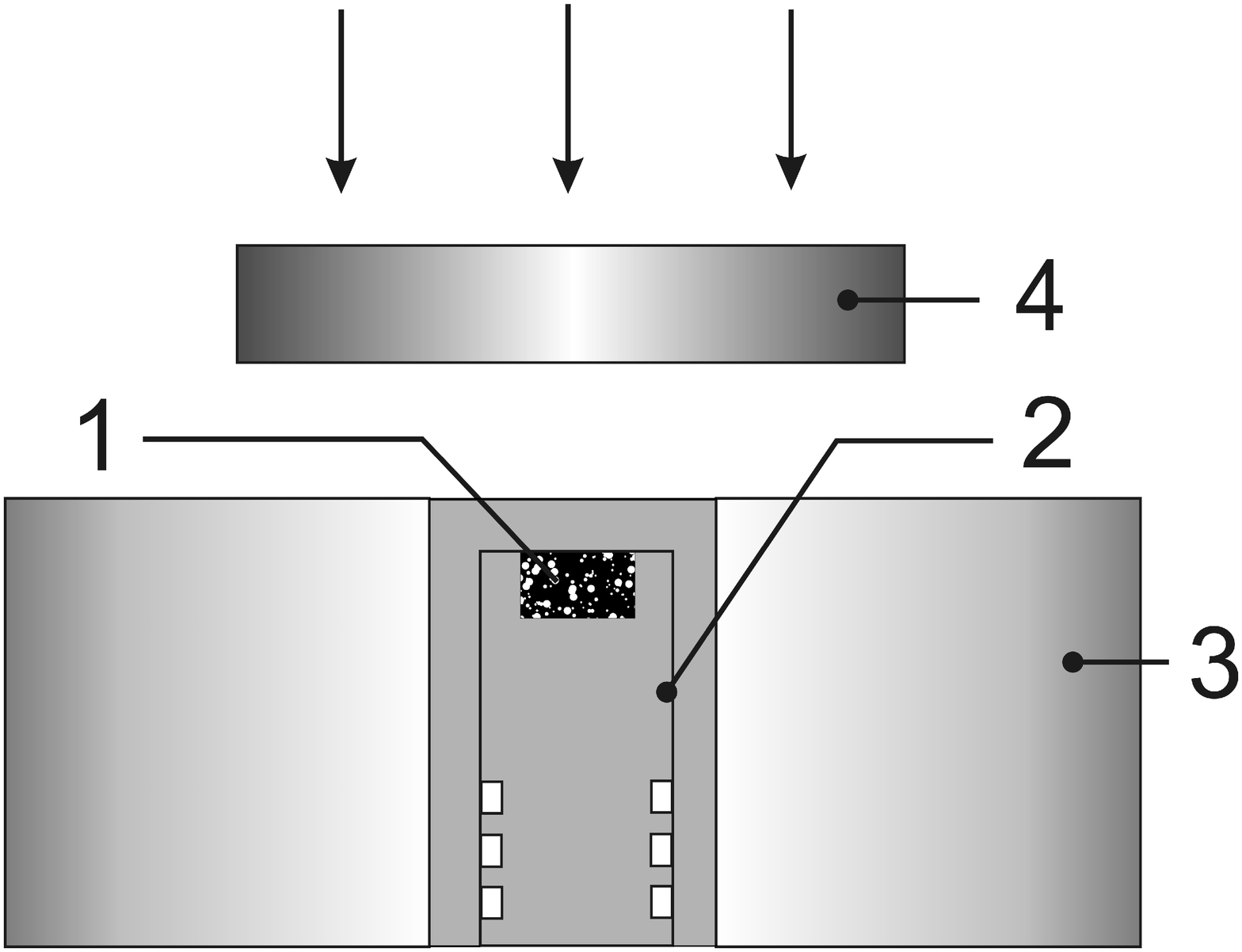}
\caption{Layout of the shock-wave pressure setup. (1) - sample, (2) -
conservation cell, (3) - guard ring, (4) - impactor. \label{Fig_1}}
\end{figure}

Layout of the shock-wave pressure setup is shown in Fig.~\ref{Fig_1}. Compacted
sample (1) reside in conservation cell (2) placed inside guard ring (3).
Aluminum impactor (4) accelerated up to 1.05\,km/s by explosion products hits
the cell lid. During the impact, the pressure is monitored by manganin pressure
gauges, whereas the average temperature of the sample is estimated by
$P$--$V$--$T$ state equations of MgO and Mg.\cite{Dorogok,ficp}

Within microseconds, the impact generates in the sample a series of consecutive
planar shock-waves of pressure.\cite{Ossipyan} Due to energy dissipation
processes in the shock-wave, the sample is heated. The heating is controlled by
adjusting the shock-wave setup. For preparing the superconducting
Mg/MgO-samples, the optimum value of the shock-wave pressure, within 1\,Mbar
range, was found 170\,kbar, which corresponds to an upper estimate of 340\,K
for the average temperature of the sample.

\begin{figure}[h]
\includegraphics[width=0.53\linewidth, angle=90]{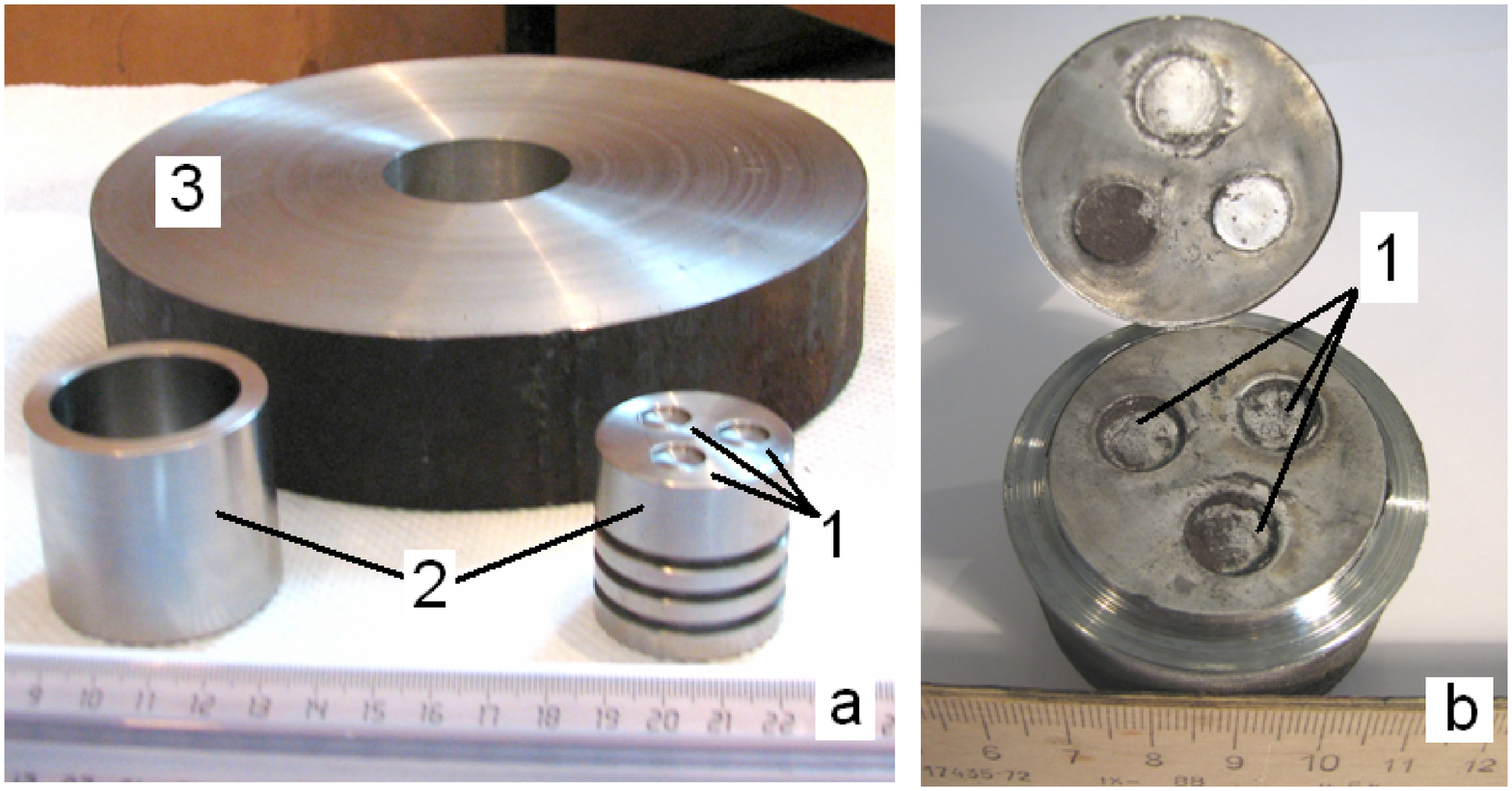}
\caption{Photographs of the unassembled virgin shock-wave pressure setup (a)
and cut-open conservation cell after the shock-wave impact (samples extracted)
(b). (1) - sample wells, (2) - conservation cell, (3) - guard ring.
\label{Fig_2}}
\end{figure}

After the shock-wave pressure treatment, the conservation cell was cut open and
the samples were extracted. The photographs of the shock-wave setup before and
after the impact are shown in Fig.~\ref{Fig_2}. The extracted samples were
vacuum-encapsulated into 5\,cm-long quartz ampoules having 0.9\,mm-thick walls
and 6\,mm outer diameter, and stored in liquid nitrogen to prevent their
degradation under normal conditions.

\subsection{Measurements}

The samples were studied by measuring the dynamic magnetic susceptibility,
$\chi=\chi'-i\chi''$, using a mutual inductance ac susceptometer,\cite{Chen}
mounted inside a variable temperature insert. To avoid sample degradation, the
measurements were done without unsealing the ampoules. The ampoule with the
sample was placed inside one of the two identical pick-up coils positioned
coaxially inside a long excitation coil that provided a uniform ac drive
magnetic field at the sample position. ac voltage induced in the pick-up coils
was fed to the differential input of a two-channel lock-in amplifier capable of
segregating the in-phase,$\chi'$, and an out-of-phase, $\chi''$, components of
the ac susceptibility.The amplitude $H_{ac}$ of the driving field ranged from
0.23 to 3.78\,Oe, the driving frequency $\nu$ from 312\,Hz to 20\,kHz, and the
superimposed dc magnetic field $H_{dc}$ up to 170\,Oe. The susceptometer was
calibrated using ac susceptibility responses to the superconducting transitions
in lead and niobium samples shaped similarly to the samples in study. An
MgB$_2$ sample sealed in a similar quartz ampoule has shown the ac
susceptibility response to the superconducting transition at 39\,K, confirming
correct thermometry of the measurement setup. The static magnetic moment of the
Mg/MgO-sample sealed in the evacuated quartz ampoule was studied by SQUID
magnetometer in the temperature range 5 - 70\,K.

Crystal structure of the samples was investigated in the temperature range 80
-- 300\,K by X-ray diffraction measurements using Oxford Diffraction Gemini R
diffractometer, MoK$_{\alpha}$ radiation, equipped with a cooling system that
enables the measurements in the flow of cold nitrogen gas. For the diffraction
measurements, the sample was extracted from the quartz ampoule in the ambience
of liquid nitrogen and rapidly (within 5\,-10\,s) mounted onto the precooled
goniometer of the diffractometer. Chemical composition of the samples was
analyzed by energy dispersive X-ray microanalysis using Oxford INCA
spectrometer.

\section{Experimental results}

\begin{figure}[h]
\includegraphics[width=0.8\linewidth, angle=90]{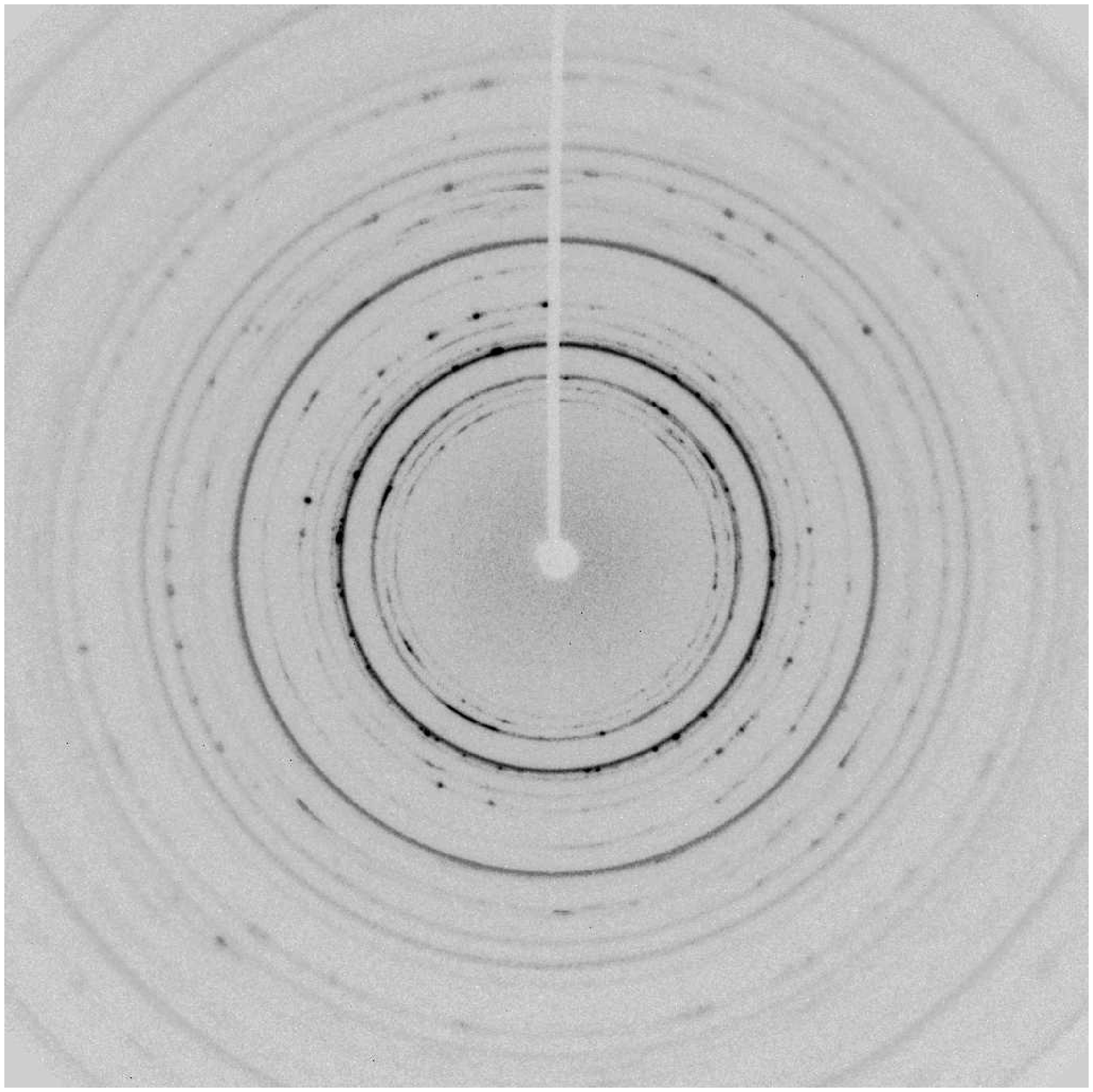}
\caption{X-ray diffraction pattern for the shock-wave pressure treated
Mg/MgO-sample in MoK$_\alpha$  radiation. The complete diffraction ring set is
a superposition of the patterns from polycrystalline MgO and Mg crystal
structures.\label{Fig_3}}
\end{figure}

Typical diffraction pattern of the shock-wave pressure treated Mg/MgO-sample
recorded at $T = 80\,$K, is shown in Fig.~\ref{Fig_3}. All the diffraction
rings in the observed pattern are a superposition of the patterns from
polycrystalline MgO ($a$ = 4.213\,\AA, space group Fm-3m) and Mg ($a$ =
3.209\,\AA, $c$ = 5.211\,\AA, space group P6$_3$/mmc) crystal
structures.\cite{diffraction}

\begin{figure}[h]
\includegraphics[width=1\linewidth, angle=0]{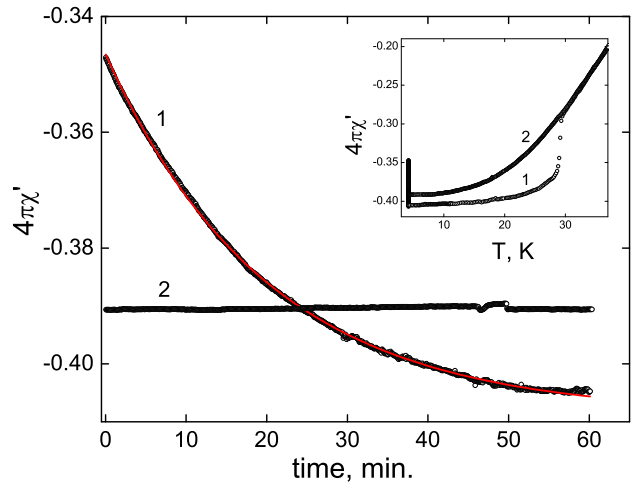}
\caption{Time evolution of $4\pi\chi'$ at $T = 4.2\,$K for the
vacuum-encapsulated shock-wave treated Mg/MgO-sample (Curve~1) and for the same
sample exposed to room temperature for 10\,hours (Curve~2). Solid curve: fit to
Curve~1 in the form $a + b\exp(-t/\tau)$, where \textit{a}, \textit{b} and
$\tau$ are fit parameters. Inset: Temperature dependencies of $4\pi\chi'$
measured before (Curve~1) and after (Curve~2) the sample exposure to room
temperature. The measurements were performed at $\nu = 5.4\,$kHz, $H_{ac} =
0.6\,$Oe, $H_{dc} = 0$.\label{Fig_4}}
\end{figure}

Each cycle of $\chi$ measurements  started from a cool-down of the
Mg/MgO-sample to 4.2\,K. Next, $H_{ac}$ was switched on to enable the
susceptibility measurements. These measurements at constant $T = 4.2\,$K have
shown that $\chi$ decays monotonously in time $t$ towards enlargement of
diamagnetism, revealing a dynamics similar to spin glass behavior.
\cite{spin_glass} Curve~1 in Fig.~\ref{Fig_4} illustrates the $4\pi\chi'(t)$
dependence measured at frequency $\nu = 5.4\,$kHz at $H_{ac} = 0.6\,$Oe and
$H_{dc} = 0$. $4\pi\chi'(t)$ was found to relax exponentially to a ground
diamagnetic state as $a + b\exp(-t/\tau)$ (shown in Fig.~\ref{Fig_4} by a solid
line) where \textit{a}, \textit{b} and $\tau$ are fit parameters. The fits to
$\chi'(t)$ curves measured at frequencies from 0.312 to 20\,kHz, give the same
time constant $\tau=19.9\,$min to within 1\%.

After a one-hour ($\sim3\tau$) delay at $T = 4.2\,$K, $\chi'(T)$ was measured
upon heating at the rate of 1-1.5\,K/min.  No visible change of the measurement
result was found with a slower heating rate. The measured $4\pi\chi'(T)$ is
shown by Curve~1 in the Inset of Fig.~\ref{Fig_4}. The drop in $4\pi\chi'(T)$
at 4.2~K corresponds to the decay of $\chi'(t)$, Curve~1 in Fig.~\ref{Fig_4}.
As the temperature increases, a step-like rise of $\chi'(T)$ is observed at
$T_{c} = 29\,$K, signifying a phase transition in the sample at this
temperature. In order to exclude the influence of the $\chi(t)$ relaxation
processes on the $\chi(T)$ measurements, all subsequent $\chi(T)$ measurements
were performed according to the described measurement cycle.

\begin{figure}[t]
\includegraphics[width=1.02\linewidth, angle=0]{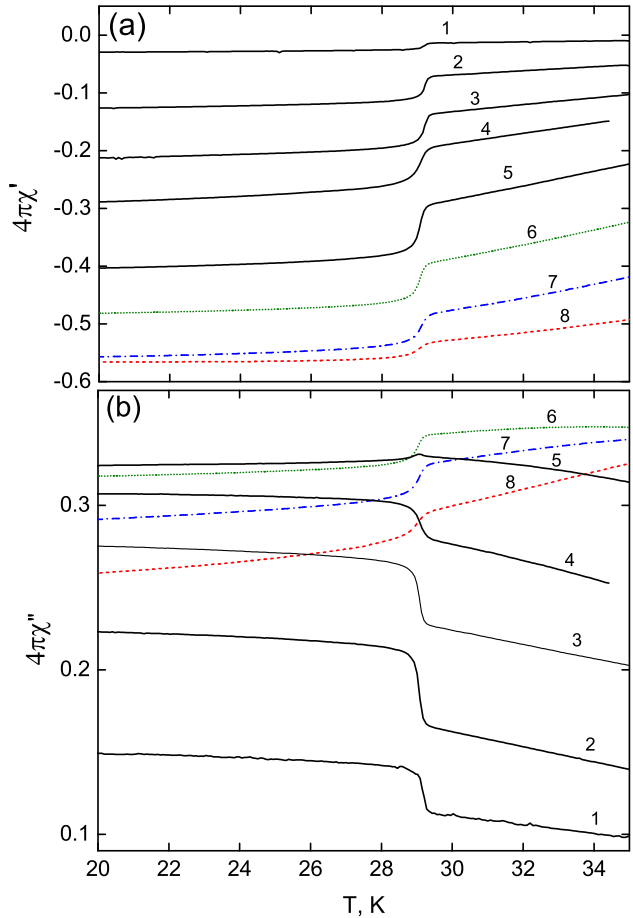}
\caption{Temperature dependencies of the real (a) and imaginary (b) parts of
$4\pi\chi$ for the Mg/MgO-sample measured at frequencies 0.312, 0.923, 1.53,
3.12, 5.4, 7.69, 10 and 20\,kHz (Curves 1 to 8, respectively). $H_{ac} =
0.6\,$Oe, $H_{dc} = 0$.\label{Fig_5}}
\end{figure}

Fig.~\ref{Fig_5} shows the temperature dependencies of $4\pi\chi'$ and
$4\pi\chi''$ measured at frequencies from 0.312 to 20\,kHz. One can see in
Fig.~\ref{Fig_5} that $T_{c}$ is essentially frequency-independent, unlike the
shapes of the $4\pi\chi'(T)$ and $4\pi\chi''(T)$ responses to the phase
transition, which change dramatically with the frequency.

\begin{figure}[t]
\includegraphics[width=1.02\linewidth, angle=0]{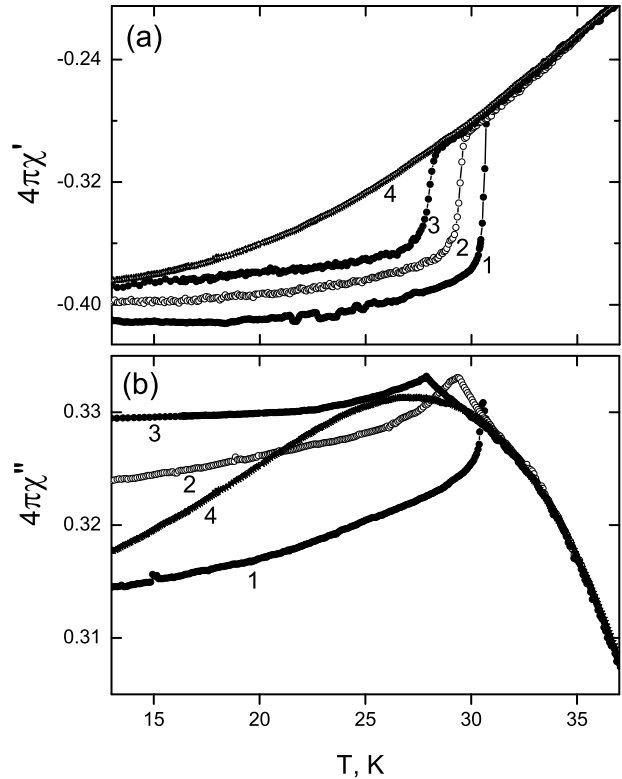}
\caption{Temperature dependencies of the real (a) and the imaginary (b) parts
of $4\pi\chi$ for the Mg/MgO-sample measured under the dc magnetic field
$H_{dc} = 0$, 84\,Oe and 168\,Oe (Curves 1 to 3, respectively). Curves~4
correspond to the same sample exposed for 10 hours to room temperature. The
measurements were performed at $\nu = 5.4\,$kHz in $H_{ac} =
0.23\,$Oe.\label{Fig_6}}
\end{figure}

\begin{figure}[t]
\includegraphics[width=1.02\linewidth, angle=0]{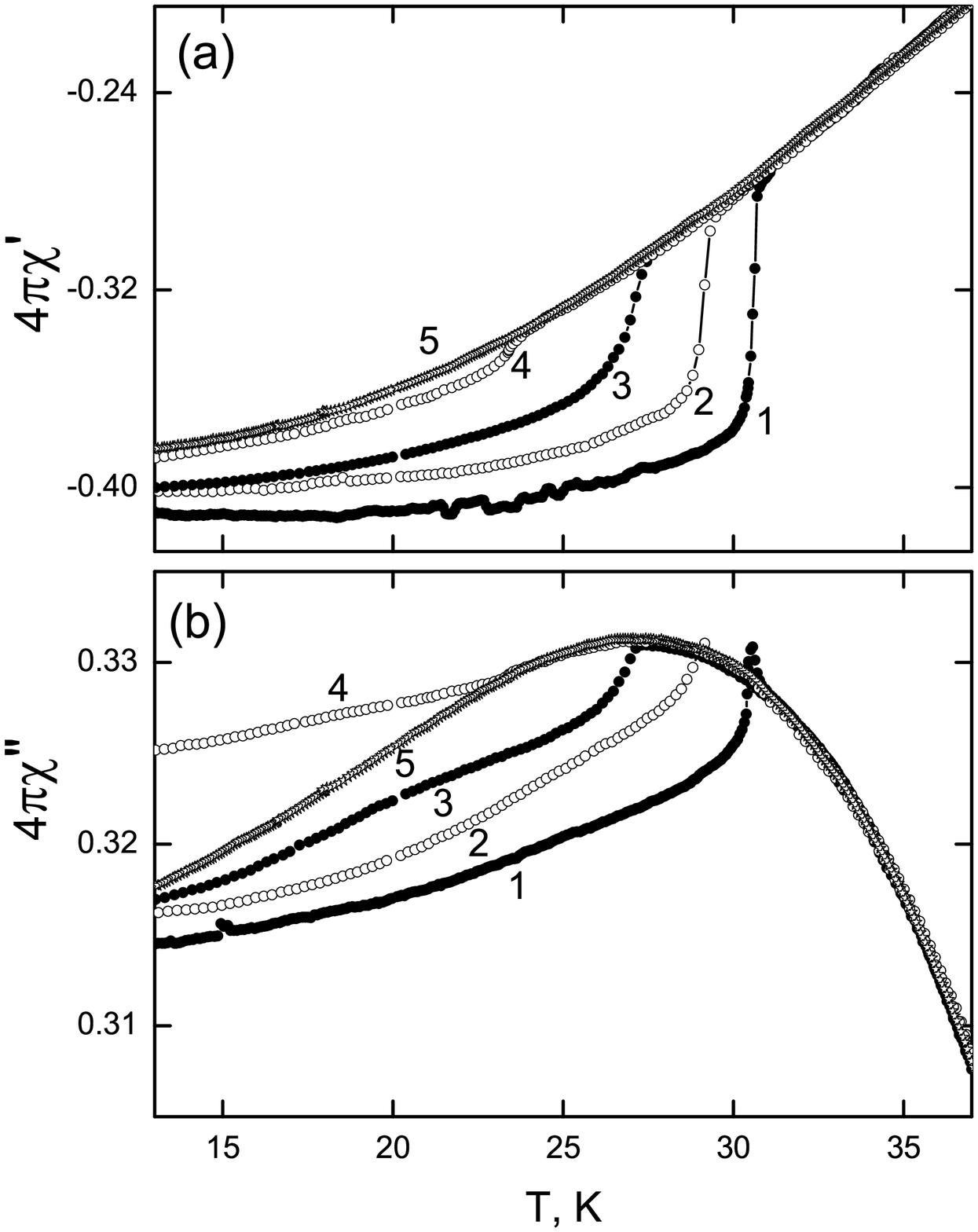}
\caption{Temperature dependencies of the real (a) and the imaginary (b) parts
of $4\pi\chi$ for the Mg/MgO-sample measured with the amplitude of the ac
driving field $H_{ac} = 0.23$, 0.6\,Oe, 1.5\,Oe and 3.78\,Oe (Curves 1 to 4,
respectively). Curves~5 correspond to the same sample exposed for 10\,h to room
temperature. The measurements were performed at $\nu = 5.4\,$kHz in $H_{dc} =
0$.\label{Fig_7}}
\end{figure}

Temperature dependencies of $\chi$ measured in a static magnetic field,
$H_{dc}$, are shown in Fig.~\ref{Fig_6}. According to Curves~1-3, an increase
in $H_{dc}$ suppresses the anomaly in $\chi(T)$. Increasing the driving ac
field amplitude, $H_{ac}$, gives a similar effect as shown in Fig.~\ref{Fig_7}.

In contrast to the ac susceptibility, the dc magnetization of the shock-wave
pressure treated Mg/MgO-samples measured using a SQUID magnetometer in a
standard zero-field-cooled and field-cooling regimes revealed no sign of the
magnetic anomaly. At temperatures from 5 to 70\,K, the sample was found
non-magnetic, demonstrating a nearly temperature-independent magnetic moment
$m\sim 10^{-6}-10^{-5}\,$e.m.u. in fields $H_{dc} = 30-300\,$Oe, respectively.

After completion of the ac susceptibility measurement series described above,
the insert with the Mg/MgO-sample was finally removed from the cryostat and
kept for $\approx10\,$h at room temperature. The following $\chi$ measurements
have shown, first of all, no time dependence of $\chi'(t)$ at 4.2\,K, see
Curve~2 in Fig.~\ref{Fig_4}. Furthermore the $T$-dependence of $\chi'$ of the
room-temperature-matured sample does not exhibit a step-like anomaly (see
Curve~2 in the Inset of Fig.~\ref{Fig_4}) and is independent on $H_{ac}$ and
$H_{dc}$, within the ranges used previously. Moreover, now $4\pi\chi'(T)$ and
$4\pi\chi''(T)$ (Curve~2 in the Inset of Fig.~\ref{Fig_4}, Curves~4 in
Fig.~\ref{Fig_6} and Curves~5 in Fig.~\ref{Fig_7}) practically coincide with
those of pure magnesium sample subjected to the shock-wave pressure treatment
under the same conditions. This denotes that the phase, which appeared in the
Mg/MgO-sample during the shock-wave pressure treatment and which is responsible
for the described anomalies in $\chi(T)$, is unstable at room temperature.

\section{Discussion}

\subsection{Evidence of metastable superconductivity}

We first focus on the sharp drop in $\chi'(T)$ observed in the shock-wave
pressure treated Mg/MgO-sample at $T_{c} = 29\,$K (Curve~1 in the Inset of
Fig.~\ref{Fig_4}), accompanied by a sharp anomaly in $4\pi\chi''(T)$
(Fig.~\ref{Fig_5}). The response to the ac magnetic field is a function of the
skin depth, $\delta\propto 1/\sqrt{\sigma\mu\nu}$, where $\sigma$ is the
conductivity and $\mu$ is the magnetic permeability of the material. According
to the dc magnetization measurements, the Mg/MgO-samples are nonmagnetic
($\mu\sim 1$) showing no anomalies in the range 5-70\,K. Therefore, the only
reason for the observed change in $\chi'(T)$ is the temperature dependence of
the electric conductivity of the sample. The observed drop in $\chi'(T)$
denotes that the conductivity rises steeply at $T_{c}\approx29\,$K as the
temperature decreases, and we suppose that this is a superconducting
transition.

This assumption is supported by the observed field dependencies of the anomaly
in $\chi(T)$. The $\chi(T)$ measurements in a static magnetic field, $H_{dc}$,
demonstrated the suppression of the $\chi(T)$ anomaly observed at
$T_{c}\approx30\,$K, $H_{ac}=0.23\,$Oe, $H_{dc}=0\,$Oe, both in size and in the
transition temperature, with increasing $H_{dc}$ (Curves 1-3 in
Fig.~\ref{Fig_6}). Increasing the amplitude of the excitation field, $H_{ac}$,
gives a similar effect, see Curves 1-4 in Fig.~\ref{Fig_7}. The $\chi'(T)$ and
$\chi''(T)$ dependencies presented in Figs.~\ref{Fig_6} and \ref{Fig_7} are
typical for superconducting materials \cite{Hein,Ishida} which implies a
superconducting transition in the Mg/MgO sample.

None of the constituents of the sample, neither metallic Mg nor MgO are
superconductors in the bulk. Moreover, none of them taken separately and
subjected to the shock-wave pressure treatment under the same conditions,
demonstrate any anomaly in $\chi(T)$. We believe therefore that the
superconductivity in the shock-wave treated Mg/MgO sample is related to the
interfacial layer formed between MgO and metallic Mg phases.

According to the results of both the ac susceptibility and the dc magnetization
measurements, the superconducting interface in the Mg/MgO-sample does not form
a closed superconducting surface capable of trapping the magnetic flux and
keeping it constant, but rather consists of weakly linked granular
superconducting two-dimensional islands.

First of all, that explains why the superconducting response is not detectable
in the zero-field-cooled regime of the dc magnetization measurement, owing to
the rapid decay of the surface shielding current after applying the dc magnetic
field. On the contrary, in the ac mode the alternating magnetic field maintains
the shielding current enabling to detect superconductivity even in
discontinuous, energy dissipative superconducting loops formed by small,
compared to the London penetration depth, weakly linked superconducting
clusters.

Secondly, at temperatures well below $T_{c}$, $4\pi\chi'(T)>-1$ and $\chi''(T)>
0$ (Figs.~\ref{Fig_5} - \ref{Fig_7}) which indicates an imperfect ac shielding,
because a closed superconducting surface assumes $4\pi\chi'(T)=-1$ and
$\chi''(T)= 0$. Non-zero $\chi''$ at $T\ll T_c$ means that the
superconductivity in the Mg/MgO-sample is accompanied by energy dissipation,
which may take place in a granular superconducting structure due to inter- and
intra-grain flux motions. Besides, systematic Y-axis shift of the $\chi''(T)$
with frequency (Fig.~\ref{Fig_5}\,(b)) reveals a considerable normal current
contribution to the energy dissipation due to partial penetration of the ac
magnetic field into the metallic magnesium core of the sample.

We believe therefore that the superconductivity in the Mg/MgO interfacial layer
is formed by thin, compared to the London penetration depth $\lambda$,
superconducting islands embedded into a non-superconducting host matrix and
weakly coupled to each other by the proximity effect (if the host is metallic
magnesium) or Josephson tunnelling (for the MgO host).

\begin{figure}[b]
\includegraphics[width=0.55\linewidth, angle=0]{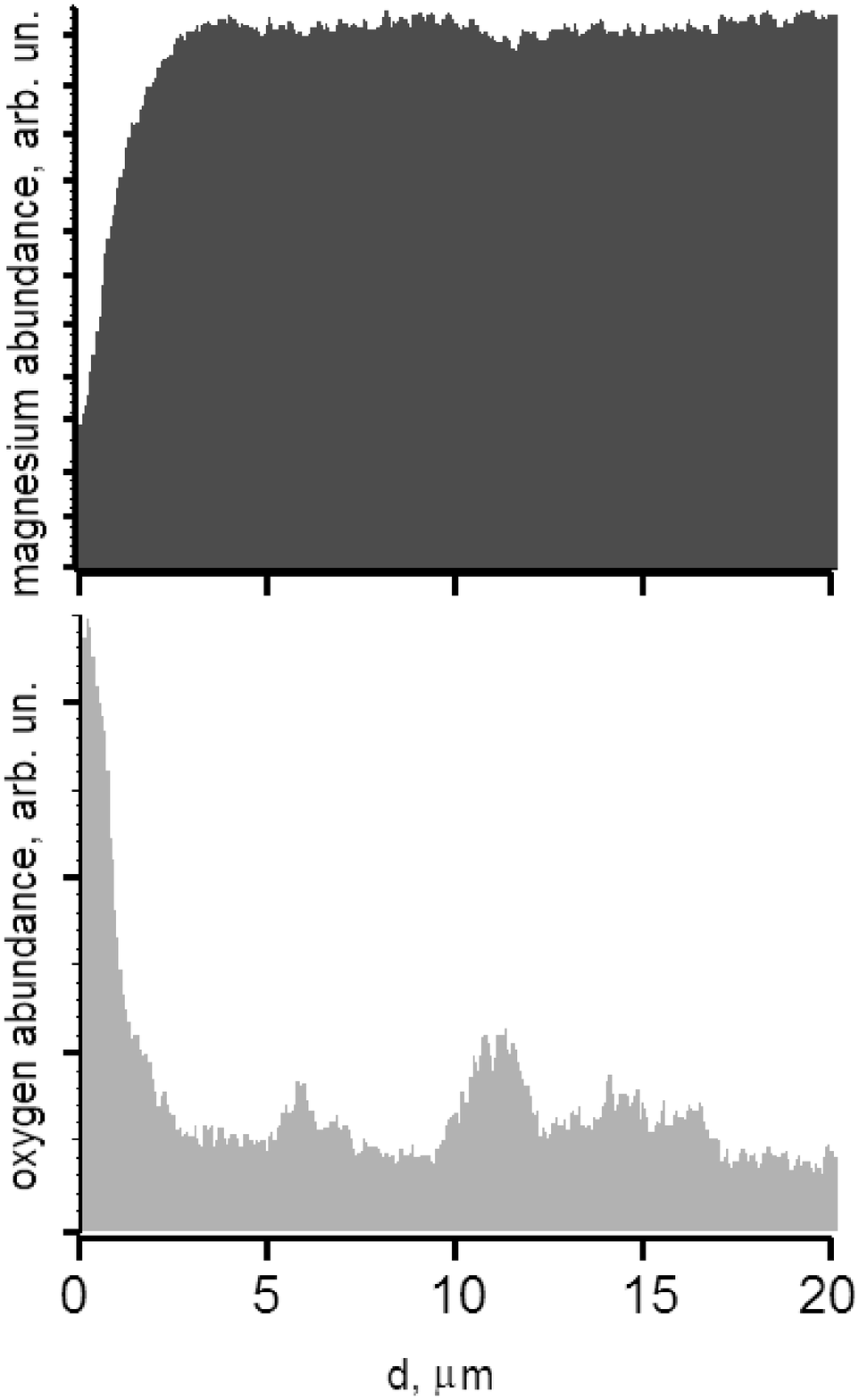}
\caption{Depth distributions of magnesium and oxygen contents in the shock-wave
pressure treated Mg/MgO-sample. Substantial change of magnesium and oxygen
concentrations is observed in the range of $\sim2\,\mu$m from the sample
surface ($d=0$) demonstrating the interfacial Mg/MgO layer depth.
\label{Fig_8}}
\end{figure}

A typical elemental distribution at the Mg/MgO-interface is shown in
Fig.~\ref{Fig_8}. Substantial change of magnesium and oxygen concentrations is
observed in the range of $\sim2\,\mu$m, representing the thickness of the
interface layer where the superconductivity occurs.

As it was mentioned above, the superconductivity of the Mg/MgO-sample is
unstable at room temperature. The instability manifests itself as a decay of
the superconducting transition after a 10\,-hour exposure of the
vacuum-encapsulated sample to room temperature (Figs.~\ref{Fig_6} and
\ref{Fig_7}), as well as vanishing of the time relaxation process of the ac
susceptibility at 4.2\,K (Fig.~\ref{Fig_4}). The reason for such instability is
apparently the oxygen/magnesium ionic diffusion processes activated at room
temperature in the Mg/MgO interface, which destroy the superconducting
interface.

The $T$-dependencies of $4\pi\chi'$ and $4\pi\chi''$ measured on the degraded
Mg/MgO-sample (Curves~4 in Fig.~\ref{Fig_6} and Curves~5 in Fig.~\ref{Fig_7})
are typical for normal nonmagnetic metals. Namely, at room temperature the skin
depth, $\delta$, of the ac magnetic field at 5.4\,kHz in magnesium is
$\approx$\,1.4\,mm, which is bigger than the sample thickness
$d\approx0.9\,$mm. As the temperature decreases, the electric conductivity of
the sample increases monotonically that leads to a decrease in $\delta$, thus
to a monotonic decrease in $\chi'(T)$. \cite{Landau} At low temperatures, the
electric conductivity of the sample becomes more flat in temperature, and
$\chi'(T)$ sets constant.

In turn, $\chi''$ characterizes the energy dissipation of the ac magnetic field
in the sample, which is also a function of $\delta$. The energy dissipation is
low at $\delta\ll r$ (high-conductivity metal) and $\delta\gg r$ (insulator),
where $r$ is the sample depth. In the intermediate region $\chi''(T)$ has a
maximum at $\delta\sim r$. \cite{Landau} This broad maximum is seen on Curve~4
in Fig.~\ref{Fig_6}\,(b) and Curve~5 in Fig.~\ref{Fig_7}\,(b) at $T\approx
27\,$K.

\subsection{Superconductive glass}

In a granular superconductor, positional disorder of individual superconducting
grains introduces randomness and frustration into the system, resulting in a
glassy behavior of the system.\cite{Ebner,Blatter} In our experiments, the
glassy behavior of the Mg/MgO-sample is evident from the observed slow dynamics
of $\chi'(t)$ relaxation into the lower-energy diamagnetic ground state at
4.2\,K (Fig.~\ref{Fig_4}). The relaxation time, $\tau = 19.9\,$min, was found
frequency-independent in the range 312\,Hz -- 20\,kHz. The insensibility of
$\tau$ to the excitation frequency is presumably a consequence of a significant
time-scale mismatch between the $\chi'(t)$ relaxation process and the ac
magnetic field oscillation period, $\tau\gg 1/2\pi\nu$: the field oscillations
are too fast compared to $\tau$ to significantly influence the $\chi'(t)$
process.

The origin of such a long $\chi'(t)$ relaxation process is rather puzzling. It
should be related to the granular structure of the superconducting
Mg/MgO-interface formed by thin, compared to $\lambda$, superconducting islands
weakly coupled to each other. However, no clear predictions concerning this
dynamics of $\chi'(t)$ in a weak ac magnetic field exist on the basis of
approaches involving collective creep or vortex-glass theories,\cite{Blatter}
or any of the different models including the critical-state,
superconducting-loop and superconducting-glass models.
\cite{Ishida,Ebner,Blatter,Kawa,Young}

Unlike $\tau$ and $T_c$, the value of the ac susceptibility is strongly
dependent on the frequency of the drive field. As one can see in
Fig.~\ref{Fig_5}\,(a), the $4\pi\chi'(T)$ plot shifts towards stronger
diamagnetism with increasing the frequency, denoting a considerable
contribution of the normal current component to $\chi'$ even at $T<T_{c}$. This
contribution is presumably related to weak links such as inter-grain boundaries
separating the superconducting grains, as well as to normal currents induced in
the metallic core of the sample due to partial penetration of the ac field.

As one can see in Fig.~\ref{Fig_5}\,(a), the step in $\chi'(T)$ at 29\,K
becomes bigger as the driving frequency increases from 0.312 to 5.4\,kHz
(Curves 1-5) and diminishes with further frequency increase (Curves 6-8). This
can be explained assuming a superposition of normal and superconducting
shielding currents contributing to $\chi'(T)$ in the granular superconductor.

Consider first the $\chi'(T)$ response at 29\,K diminishing with the frequency
($\nu\geq5.4$\,kHz, Curves 5-8 in Fig.~\ref{Fig_5}\,(a)). This effect is well
known for  bulk metallic superconductors: in a normal metal, increasing the ac
magnetic field frequency leads to diminishing skin depth $\delta$, hence a
bigger diamagnetic signal. As a result, at high frequency the normal metal just
above $T_c$ is nearly as diamagnetic as the superconductor just below $T_c$
which makes the superconducting transition indistinguishable in $\chi'(T)$.

On the other hand, the suggested granular structure of the superconducting
Mg/MgO-interface assumes intergranular weak links which can be considered as
electric capacitors connected in series into the shielding supercurrent loops.
Obviously, the higher is frequency, the better the superconducting islands are
linked, hence the stronger is the superconducting shielding. Apparently, this
explains the growth of the $\chi'(T)$ response at 29\,K at frequencies up to
5.4\,kHz (Curves 1-5 in Fig.~\ref{Fig_5}\,(a)). The two competing contributions
superimpose in the studied  Mg/MgO-sample to give a maximum of the $\chi'(T)$
response to the superconducting transition at $\nu_0 = 5.4\,$kHz.

The evolution of $\chi''(T)$ with frequency is even more complicated, see
Fig.~\ref{Fig_5}\,(b). At frequencies lower than $\nu_0 = 5.4\,$kHz (Curves
1-4) $\chi''(T)$ exhibits a step-like rise at cooling below $T_{c}\approx
29$\,K, while at higher frequencies (Curves 6-8) a step-like drop is observed
instead. At the crossover frequency $\nu_0$ (Curve~5) a cusp-like anomaly is
observed at $T_c$.

A possible explanation to such frequency behavior of $\chi''(T)$ which reflects
energy dissipation in the system, is vortex dynamics influenced by the ac drive
field. \cite{Blatter} We assume that the crossover frequency $\nu_0 = 5.4\,$kHz
matches the vortex system relaxation rate $1/\tau_0$, $2\pi\nu_0\tau_0 = 1$. At
low frequencies $\nu\ll\nu_0$, the ac field oscillations are slow compared to
the vortex relaxation time $\tau_0$, which enables the vortices respond to
variation of the field. This gives rise to energy dissipative vortex motion,
reflected in a step-like increase of $\chi''(T)$ upon cooling through
$T_c\approx 29\,$K (Curves 1-4 in Fig.~\ref{Fig_5}\,(b)). At high frequencies
$\nu\gg\nu_{0}$, the drive field oscillates too fast to disturb the vortex
system which disables the energy dissipation through vortex motion. As a
result, a step-like decrease of $\chi''(T)$ is observed upon cooling through
$T_c\approx 29\,$K (Curves 6-8 in Fig.~\ref{Fig_5}\,(b)), which reflects
shielding of the energy-dissipative interior of the sample by the
supercurrents.

\subsection{Nature of the  Observed  Superconductivity}

None of the constituents of the sample, neither metallic Mg nor MgO are
superconductors in the bulk. Moreover, none of them taken separately and
subjected to the shock-wave pressure treatment under the same conditions,
demonstrate any anomaly in $\chi(T)$. We believe therefore that the
superconductivity in the shock-wave treated Mg/MgO sample is related to the
interfacial layer formed between MgO and metallic Mg phases.

Unlike the $\chi(T)$ dependencies that change dramatically after exposing the
sample to room temperature due to decay of the superconducting fraction, no
change has been detected in the X-ray diffraction patterns of the Mg/MgO-sample
after the room-temperature exposure. The fraction of the superconducting phase
is therefore too small to be detected by the X-ray diffraction technique.

One may reasonably suggest ascribing the superconductivity in the Mg/MgO sample
to some unknown superconducting magnesium oxide phase(s), unstable under normal
conditions. However, this phenomenon has been discovered for other similar
objects based on metals of various groups and their oxides (Na/NaO$_x$,
Cu/CuO$_x$, Al/Al$_2$O$_3$ and Fe/FeO$_x$). \cite{Na,Cu,Al,Fe} This indicates
the generality of the observed phenomenon, rather than a unique property of
Mg/MgO-system.

Alternatively, consider nanometer-sized metallic clusters which may
spontaneously arise in the metal-oxide interface layer. Delocalized electrons
of such cluster form energy shells similar to those in atoms or nuclei. Under
specific conditions providing narrow, partially filled energy band at the Fermi
level, superconducting pairing in such objects is expected to become very
strong, forming a new hypothetical family of high-$T_{c}$ ($>$150\,K)
superconductors. \cite{Kresin}

Besides, superconducting two-dimensional metal/oxide interfacial regions of
sub-micron size may spontaneously arise in a relatively thick ($\sim 1\,\mu$m)
boundary layer separating the metal and its oxide. In such interfaces which
play a role of asymmetric confining potentials in the system of free electrons,
the lack of spatial inversion symmetry may result in a topological change of
the Fermi surface due to the spin-orbit splitting, \cite{Cappelluti} and lead
to the enhanced superconductivity.

\section{Acknowledgements}

We are grateful to V.~V.~Ryasanov, V.~F.~Gantmakher and G.~M.~Eliashberg for
useful discussions of the results. The work has been supported by RAS Presidium
Programs "Quantum physics of condensed matter" and "Thermal physics and
mechanics of extreme energy impacts and physics of strongly compressed matter",
as well as RFBR 13-02-01217a - project.

\end{document}